# CORRELATION BETWEEN STRUCTURE AND C-AFM CONTRAST OF 180-DEGREE DOMAIN WALLS IN RHOMBOHEDRAL BaTi0$_3$

Eugene A. Eliseev[1], Peter V. Yudin[2], Sergei V. Kalinin[3], Nava Setter[2], Alexander K. Tagantsev[2] and Anna N. Morozovska[4,*]

[1] Institute for Problems of Materials Science, National Academy of Sciences of Ukraine, 3, Krjijanovskogo, 03142 Kiev, Ukraine

[2] Ceramics Laboratory, Swiss Federal Institute of Technology (EPFL), CH-1015 Lausanne, Switzerland

[3] Center for Nanophase Materials Science, Oak Ridge National Laboratory, Oak Ridge, Tennessee, 37831, USA

[4] Institute of Physics, National Academy of Sciences of Ukraine, 46, pr. Nauki, 03028 Kiev, Ukraine

## Abstract

Using Landau-Ginzburg-Devonshire theory we describe 180-degree domain wall structure, intrinsic energy and carrier accumulation in rhombohedral phase of BaTiO$_3$ as a function of the wall orientation and flexoelectric coupling strength. Two types of domain wall structures (phases of the wall) exist depending on the wall orientation. The low-energy "achiral" phase occurs in the vicinity of the {110} wall orientation and has odd polarization profile invariant with respect to inversion about the wall center. The second "chiral" phase occurs around {211} wall orientations and corresponds to mixed parity domain walls that may be of left-handed or right-handed chirality. The transformation between the phases is abrupt, accompanied with 20-30% change of the domain wall thickness and can happen at fixed wall orientation with temperature change. We suggest that the phase transition may be detected through domain wall thickness change or by c-AFM. The structure of the domain wall is correlated to its conductivity through polarization component normal to the domain wall, which causes free carriers accumulation. Depending on the temperature and flexoelectric coupling strength relative conductivity of the wall becomes at least one order of magnitude higher than in the single-domain region, creating c-AFM contrast enhancement pronounced and detectable.

[*] Corresponding author e-mail anna.n.morozovska@gmail.com

# 1. INTRODUCTION

There is a renewed interest to the internal structure and conductivity of domain walls (DW) in ferroelectrics. Polarization behavior in the DW in multiaxial ferroelectrics has been studied since the early days of ferroelectricity [1, 2, 3] till nowadays [4, 5, 6, 7, 8, 9]. Continuum Landau-Ginzburg-Devonshire (LGD) theory appeared relevant and powerful for modeling DWs polar structure, energy and electronic properties of uniaxial [10, 11, 12] and multiaxial ferroelectrics [13, 14, 15, 16, 17], and incipient ferroelectrics – ferroelastics [18]. Despite the fact that intrinsic width of DWs in multiaxial ferroelectrics can be about several lattice constants, domain wall structure calculated using LGD is in quantitative agreement with the one calculated by density functional theory (DFT) [19, 20, 21].

The DW conductivity mechanism stemming from the screening of the potential jump caused by depolarization field [3, 22, 23 24] was justified by numerous studies. Recent studies have proven that nominally uncharged DWs [25, 26, 27, 28] and vortex structures [29] in BiFeO$_3$ exhibit strongly enhanced room-temperature current-AFM (**c-AFM**) contrast in comparison with a single-domain regions. Nominally uncharged 180-degree DWs [30] and nanodomains [31] in Pb(Zr,Ti)O$_3$ appeared indeed conductive.

In general case polarization vector inside a DW can have all three components. The first one parallel to the spontaneous polarization $\pm P_S$ in the domains is regarded as the Ising component; the second one, also parallel to the wall plane, but perpendicular to Ising component, which vanishes far from the wall, is regarded as Bloch component; and the third one normal to the wall is regarded as Neel component (see **Figure 1a**). According to recent theoretical studies mixed Ising-Bloch-Neel 180-degree walls are present in a wide range of ferroelectric materials. Using LGD and DFT, Lee et al [32] and Rakesh et al [33] reported about their presence in PbTiO$_3$, LiNbO$_3$ and thin strained films of **BaTi0$_3$** (BTO), at that the Neel component appeared comparable with the Ising one

Studies by Marton, Hlinka et al [13, 14, 15, 19] reveal great interest to the rhombohedral phase of BTO, where original internal structure and behaviour of neutral domain walls are discovered. Domain walls with large Bloch component, commensurate with Ising component are predicted [14]. Neutral 180-degree DW having {211} orientation are reported to undergo phase transition from Ising-Bloch to purely Ising state [15] under application of moderate stress. These results for rhombohedral BTO are consistent with DFT – calculations [21].

However in these works [13-15, 21, 32, 33] only several wall orientations were considered. Questions about energetically preferable orientations and about guidelines for the experimental observation of the phase transition in the wall still remain open. Besides, the impact of the flexoelectric coupling on the DW structure was not taken into account. Flexoelectric effect describes the coupling of polarization with strain gradient and polarization gradient with the strain [34, 35, 36, 37]. Flexoelectricity-related electromechanical coupling is shown to be strong by LGD in conventional ferroelectrics [38, 39, 40, 41, 42, 43] and this is confirmed by experimental trends [44, 45, 46]. Angular energy anisotropy of DW in uniaxial ferroelectrics LiNbO$_3$ and LiTaO$_3$ with trigonal symmetry (which is similar to rhombohedral BTO) was studied by Scrymgeour et al [10]. Despite Bloch-Neel polarization component appeared small (~$10^{-4}$ C/m$^2$), they lead to considerable (1-3%) hexagonal energy anisotropy of domain walls.

These motivate us to perform LGD-based study of 180-degree DW structure in the rhombohedral phase of BTO with account of flexoelectric coupling and explore angular anisotropy of DW energy. Our studies confirm the possibility of phase transitions between two different wall structures. However, we show that the case of {211}-wall orientation considered in [15] is energetically unfavorable. We show that similar phase transition may be achieved via the change of DW orientation and, for a narrow domain of orientations, via temperature change at constant orientation. We suggest possible experimental ways of observation of the phase transition in the domain wall. One of them utilizes the correlation between domain wall structure and conductivity. We show that Neel component in rhombohedral BTO lead to free charges accumulation pronounced enough to be detected by c-AFM and this is partially conditioned by flexoelectric coupling.

The structure of the article is as follows. In section 2 we describe fundamental properties of the domain walls, which are not conditioned by flexoelectric effect. The changes introduced by the flexoelectric effect are considered in section 3. In the section 4 the c-AFM contrast of the domain walls is discussed.

## 2. CLASSICAL CONSIDERATION OF 180-DEGREE DOMAIN WALL

For the sake of clarity in this section we consider the problem in the frame of simple LGD- theory neglecting flexoelectric coupling and semiconducting properties of BTO (dielectric limit). We will prove a posteriori that conductivity has negligible influence on DW structure and intrinsic energy. Additional features introduced by flexoelectric coupling are discussed in the section 3.

### 2.1. Statement of the problem

Let us consider nominally uncharged 180-degree DW in the bulk of BTO single crystal. The polarization profile in the wall can be derived within LGD theory. We base our calculations on Gibbs potential $G$ with differential $dG = E_i dP_i - u_{ij} d\sigma_{ij}$, where $E_i$ is electric field (including the depolarization one), $u_{ij}$ are elastic strains, $\sigma_{ij}$ are elastic stresses, $P_i$ are polarization components related to the soft mode. For the $m\bar{3}m$ symmetry in the crystallographic frame the expression for the Gibbs potential has the form [16, 17]:

$$G = G_{polar} + G_{grad} + G_{striction} + G_{elastic} \qquad (1)$$

Where $G_{polar} = a_i P_i^2 + a_{ij} P_i^2 P_j^2 + a_{ijk} P_i^2 P_j^2 P_k^2$ is Landau part, $G_{grad} = \dfrac{g_{ijkl}}{2} \dfrac{\partial P_i}{\partial x_j} \dfrac{\partial P_k}{\partial x_l}$ is gradient or Ginsburg part, $G_{striction} = -Q_{ijkl} \sigma_{ij} P_k P_l$ is electrostriction term, $G_{elastic} = -\dfrac{s_{ijkl}}{2} \sigma_{ij} \sigma_{kl}$ is elastic contribution. Hereinafter $a_i$, $a_{ij}$ and $a_{ijk}$ are LGD-expansion coefficients of the 2$^{nd}$, 4$^{th}$ and 6$^{th}$ order dielectric stiffness tensors correspondingly, gradient coefficients are $g_{ijkl}$, $Q_{ijkl}$ are 4-th second rank electrostriction tensors coefficients, $s_{ijkl}$ are elastic compliances. Numerical values of the tensor components are listed in the **Table S1**, **Suppl. Mat**.

Regarding that all physical quantities can depend only on the distance $\tilde{x}_1$ from the DW plane $\tilde{x}_1 = 0$, it make sense to define them in the coordinate set $\{\tilde{x}_1, \tilde{x}_2, \tilde{x}_3\}$ rotated with respect to the cubic crystallographic axes $\{x_1, x_2, x_3\}$ as shown in **Fig. 1b**. Here α is the wall tilt angle about the cube spatial diagonal with respect to the <101> plane.

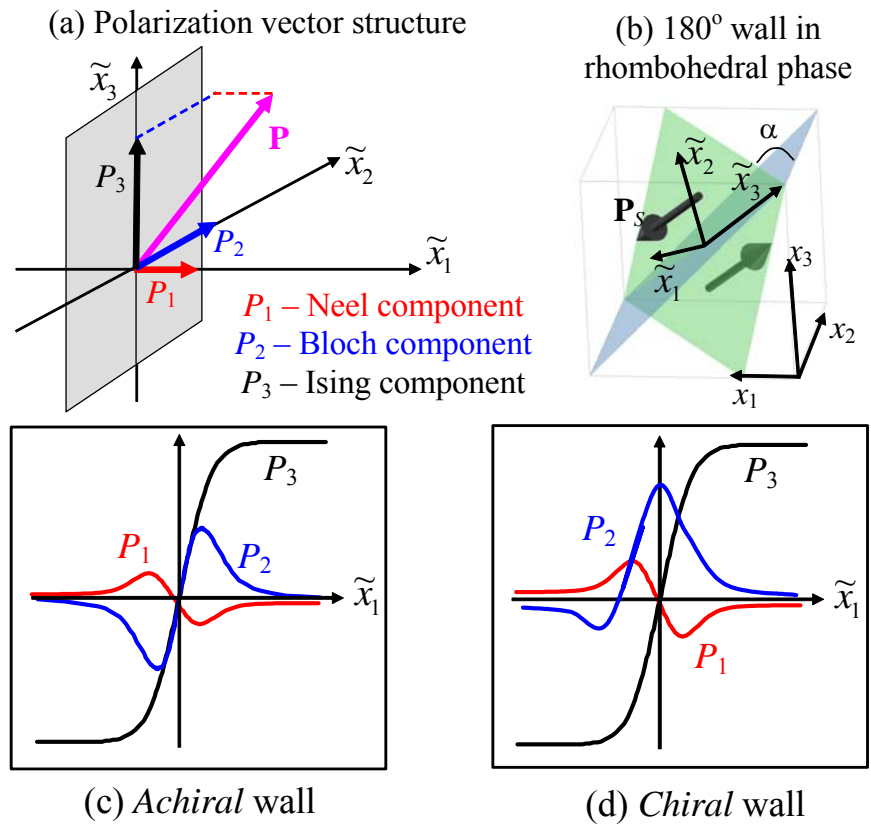

**Figure 1.** (a) Polarization vector structure. (b) Rotated coordinate frame $\{\tilde{x}_1, \tilde{x}_2, \tilde{x}_3\}$ choice for 180-degree nominally uncharged domain walls in the rhombohedral ferroelectric BTO; α is the wall tilt angle counted from crystallographic plane <101>. The distance from the wall plane is $\tilde{x}_1$. (c,d) Schematics of the polarization component distribution inside achiral and chiral domain walls.

We rewrite (1) in the new reference frame to obtain $\tilde{G}$ and to derive Euler-Lagrange equations [47] for polarization components $\tilde{P}_i$ and equations of state for elastic stresses $\tilde{\sigma}_{ij}$ correspondingly:

$$\frac{\partial \tilde{G}}{\partial \tilde{P}_i} - \frac{\partial(\partial \tilde{G})}{\partial(\partial \tilde{P}_i/\partial \tilde{x}_1)} = \tilde{E}_i \quad (2a)$$

$$\frac{\partial \tilde{G}}{\partial \tilde{\sigma}_{ij}} = -\tilde{u}_{ij} \quad (2b)$$

External field is regarded absent, so $\tilde{E}_2 = \tilde{E}_3 = 0$. The depolarization field $\tilde{E}_1^d$, caused by the inhomogeneity of $\tilde{P}_1(\tilde{x}_1)$, can be derived from Maxwell equation $\text{div}\mathbf{D} = 0$, where $\mathbf{D}$ is the electric displacement, as [10]:

$$\tilde{E}_1^d(\tilde{x}_1) \approx \frac{-\tilde{P}_1(\tilde{x}_1)}{\varepsilon_0 \varepsilon_b} \quad (3)$$

Universal dielectric constant is $\varepsilon_0 = 8.85 \times 10^{-12}$ F/m, $\varepsilon_b$ is the background dielectric permittivity unrelated with the soft mode. The boundary conditions are $\tilde{P}_3(\tilde{x}_1 \to \pm\infty) = \pm\tilde{P}_S$, $\tilde{P}_{1,2}(\tilde{x}_1 \to \pm\infty) \to 0$, $\tilde{E}_1(\tilde{x}_1 \to \pm\infty) \to 0$ and $\tilde{\sigma}_{ij}(\tilde{x}_1 \to \pm\infty) = 0$. By setting $\tilde{P}_3(\tilde{x}_1 = 0) = 0$ we determine the origin at the domain wall plane.

By solving equations of state along with mechanical equilibrium conditions $\partial \tilde{\sigma}_{1j}/\partial \tilde{x}_1 = 0$ and compatibility relation $e_{ill} e_{j1n}(\partial^2 \tilde{u}_{ln}/\partial \tilde{x}_1^2) = 0$ we eliminate mechanical variables. Explicit form of these equations and elastic stresses in the rotated coordinate frame are listed in the **Suppl. Mat., Appendixes S1-S2.** Below we present the results of numerical calculations of the Eqs.(2)-(3).

### 2.2. DW Structure

Depending on the wall orientation two types of domain walls can realize: chiral and achiral. The wall is achiral if its profile is invariant upon the inversion with respect to the wall center. In achiral wall all the components are odd functions of the $\tilde{x}_1$ - coordinate [**Fig. 1c**]. In the chiral wall type the Bloch and Neel components are of mixed $\tilde{x}_1$ - parity, i.e. contain odd and even compound [**Fig. 1d**]. As follows from the symmetry of the problem (the governing equations and boundary conditions are invariant upon the inversion with respect to the wall center), the chiral walls are bistable, i. e. the polarization may draw right or left helices on passing from one domain to the other. In the both wall types all the three components, Ising, Bloch, and Neel are present. Note that in contrast to the tetragonal symmetry [48], in the rhombohedral phase Neel polarization component is nonzero even under the absence of the flexoelectric coupling.

Distributions of Neel $\widetilde{P}_1(\widetilde{x}_1)$ and Bloch $\widetilde{P}_2(\widetilde{x}_1)$ polarization components across the 180-degree wall are shown in **Fig. 2a,b** for $\alpha = 0$, $\pi/10$, $\pi/6$ correspondingly. The Ising component $\widetilde{P}_3(\widetilde{x}_1)$, which is not illustrated, has standard kink profile, which is almost weakly α-dependent. One can see that $\widetilde{P}_1(\widetilde{x}_1)$ is about two order of magnitude smaller than $\widetilde{P}_2(\widetilde{x}_1)$ as it is suppressed by the depolarizing field $\widetilde{E}_1(\widetilde{x}_1)$. Although, as we show in section 4, despite its smallness, it has important influence on the conductivity of the DW.

The structure of the wall is strongly dependent on the DW orientation. We illustrate this dependence in **Fig. 2c-d** by plotting maximal ($\widetilde{P}_i^{\max}$) and minimal ($\widetilde{P}_i^{\min}$) values of the Bloch and Neel components, indicated in **Fig. 2a,b,** as functions of the domain wall rotation angle α. The achiral solution (dotted line in **Fig. 2c-d**) is available for any DW orientation. However for the wall orientations around $\alpha = \pi/6 + m\pi/3$ ( $m = 0, 1, 2,...$ ) it becomes metastable, because the chiral solution becomes energetically preferable (see **Fig 3** from section 2.3). The true solutions, corresponding to minimal intrinsic energy are shown by the solid and dash-dotted curves for left- and right-handed solutions respectively in **Fig. 2c-d.** Thus if we virtually rotate the DW, it undergoes a phase transition from achiral to chiral state at $\alpha \approx (2m+1)\pi/12$, $m = 0,1,2$.

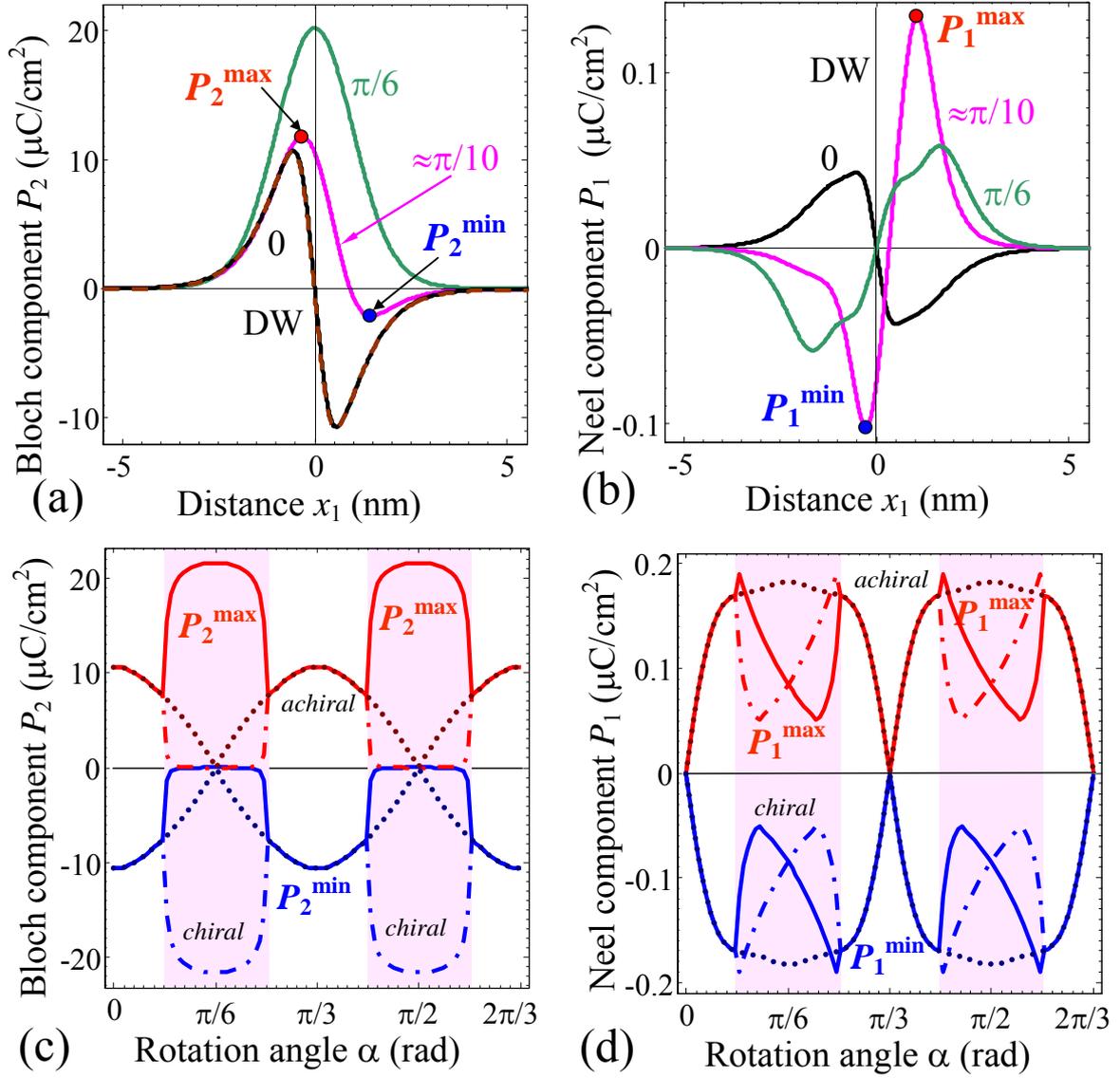

**Figure 2.** Profiles of (a) Bloch $\widetilde{P}_2(\widetilde{x}_1)$ and (b) Neel $\widetilde{P}_1(\widetilde{x}_1)$ polarization components calculated across the DW for rotation angles $\alpha = 0$, $\pi/10$, $\pi/6$ (specified near the curves), temperature 200 K and BTO parameters listed in the **Table S1**. Angular dependence of maximal (**red** upper curves labeled $\widetilde{P}_i^{\max}$) and minimal (**blue** bottom curves labeled $\widetilde{P}_i^{\min}$) values of Bloch (c) and Neel (d) polarization components. Absolutely stable solutions are shown by the solid and dash-dotted curves for left- and right-handed solutions correspondingly. **Achiral** solutions are shown by dotted curves. Filled rectangles indicate the region of absolute stability of **chiral** walls. Empty regions correspond to achiral wall absolute stability regions.

### 2.3. DW Energy

To calculate the free energy of the DW we perform Legendre transformation of the potential (1) [49] as $F = \int_{-\infty}^{\infty} \left( \widetilde{G} + \widetilde{u}_{ij} \widetilde{\sigma}_{ij} - \widetilde{P}_1 \widetilde{E}_1^d / 2 \right) d\widetilde{x}_1$. Dependencies of the DW energy on the wall orientation are shown in **Figure 3**. The obtained energy anisotropy (see polar plot **3b**) explains the anisotropic hexagon-like domains observed experimentally in BTO [50]. One can see from the **Figure 3a** that energetically preferable orientations $\alpha = m\pi/3$, $m = 0,1,2$ correspond to achiral walls. In contrast, chiral walls are realized in the vicinity of the energy maximums $\alpha = \pi/6 + m\pi/3$.

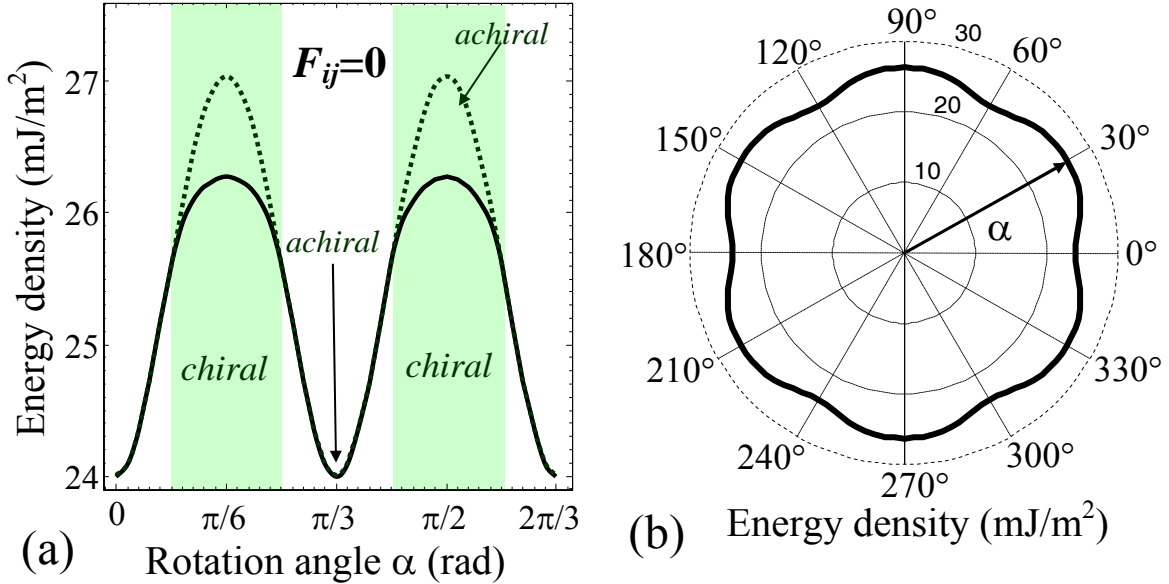

**Figure 3.** Angular dependence (a) and polar plot (b) of the 180-degree DW energy density calculated in rhombohedral phase of BTO, T=200 K. Solid curves correspond to the true solution. The energy of achiral solution is shown by dotted curves. Filled and empty rectangles indicate the regions of absolute stability of chiral and achiral solutions respectively.

Thus the phase transition mechanism proposed in [15] faces with difficulties to maintain {211} wall orientation. We reconsider the question of the phase transition with account of angular dependence of the wall structure and energy studied.

### 2.4. Phase transition in the wall

Phase transition between chiral and achiral phases inside DW in the rhombohedral BTO is originally proposed by Hlinka et al [15]. It is suggested to apply mechanical stress to switch {211} wall between the two phases. In our study we observe both chiral and achiral phases at zero stress and we are aiming to show that there is temperature-driven phase transition possible between the two phases. From the point of view of theory, the most straightforward way to pass through the transition is by virtually rotating the domain wall from achiral {110}-"ground state" to chiral state. Or one can imagine a cylindrical domain wall where the phase transition occurs in certain spatial points. Thus we start from angular dependence of chirality. As a measure of chirality we utilize the parameter $C = \int_{-\infty}^{\infty}\left(\widetilde{P}_3 \frac{d\widetilde{P}_2}{d\widetilde{x}_1} - \widetilde{P}_2 \frac{d\widetilde{P}_3}{d\widetilde{x}_1}\right)d\widetilde{x}_1$, introduced by Salje et. al. [51]. In similar manner we introduce the chiral dipole moment or "bichirality" $biC = \int_{-\infty}^{\infty}\left(\widetilde{P}_3 \frac{d\widetilde{P}_2}{d\widetilde{x}_1} - \widetilde{P}_2 \frac{d\widetilde{P}_3}{d\widetilde{x}_1}\right)\widetilde{x}_1 d\widetilde{x}_1$. Somehow the $C$ and $biC$ parameters characterize even and odd Bloch-polarization-compounds respectively.

The phase transition in the wall from achiral to chiral state is illustrated in **Fig. 4** by the dependence of $C$-parameter on the wall rotation angle $\alpha$ ("rotation-driven"). The transition happens at angles $\alpha_{cr}^m \approx \pi/6 \pm \pi/12 + m\pi/3$ for $T$=200K. The critical angles exhibit some weak temperature dependence, as shown in the **Fig. 4b** for the one $\alpha_{cr}^* \sim 5\pi/12$. Thus there is a narrow region of wall orientations $0.4\pi - 0.415\pi$, where phase transition may be achieved by temperature change at constant wall orientation, as illustrated in the **Fig. 4b**. The behaviour of Bloch and Neel components near such temperature-driven phase transition is illustrated in **Fig. 4c-d**, where noticeable jump of their maximal values is observed. The jump of the Bloch component (**Fig. 4c**) is far not small (about 2 times), and we dare to propose the way of its experimental observation through its correlation with the relatively small jump on the Neel component (**Fig. 4d**), which can be detected from c-AFM at different temperatures as it will be discussed in the section 4. The big jump on the Bloch component can lead to the non-trivial behaviour of the DW width in the vicinity of the phase space point $\{\alpha_{cr}^*, T_{cr}^*\}$, where the achiral wall becomes more stable than the chiral one. **Figure 5** demonstrate such temperature behaviour of the DW width calculated for the Ising polarization component at the level 0.5 with respect to the saturation value. Since the jump on DW width is notable (see solid and dashed curves in **Fig. 5**), the predicted temperature-induced phase-transition from chiral to achiral wall may possibly be verified experimentally from the domain wall width temperature measurements by using STEM.

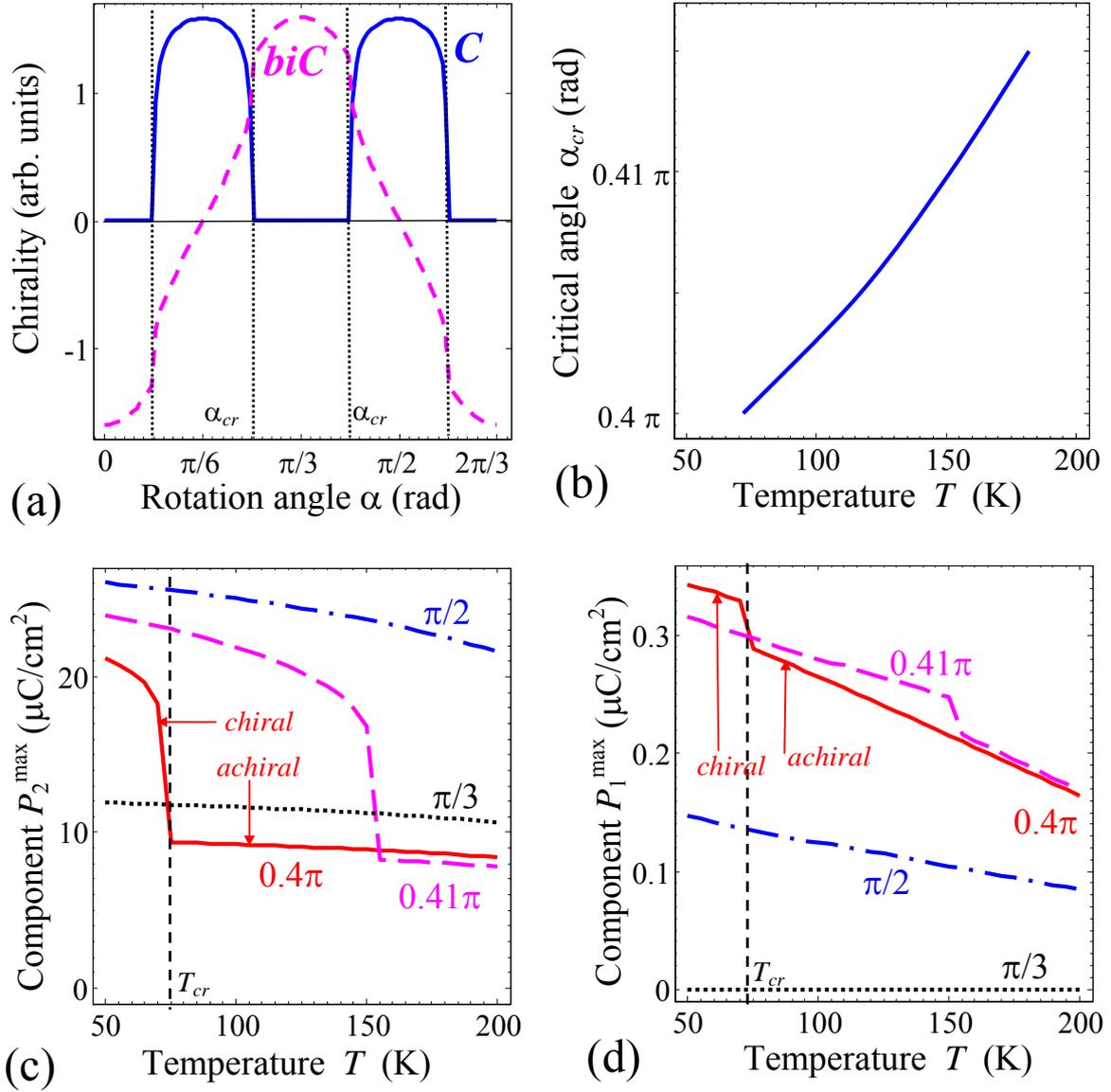

**Figure 4.** (a) Chirality $C$ (solid curve) and bichirality $biC$ (dashed curve) vs. wall rotation angle $\alpha$ calculated at 200 K. Only left-handed wall is shown. (b) Temperature dependence of maximal Bloch (c) and Neel (d) components calculated in rhombohedral ferroelectric phase of BTO for different rotation angles $\alpha = \pi/3$ (minimal energy); $\alpha = \pi/2$ (maximal energy) and $\alpha = 0.4\pi, 0.41\pi$ (phase transition) specified near the curves.

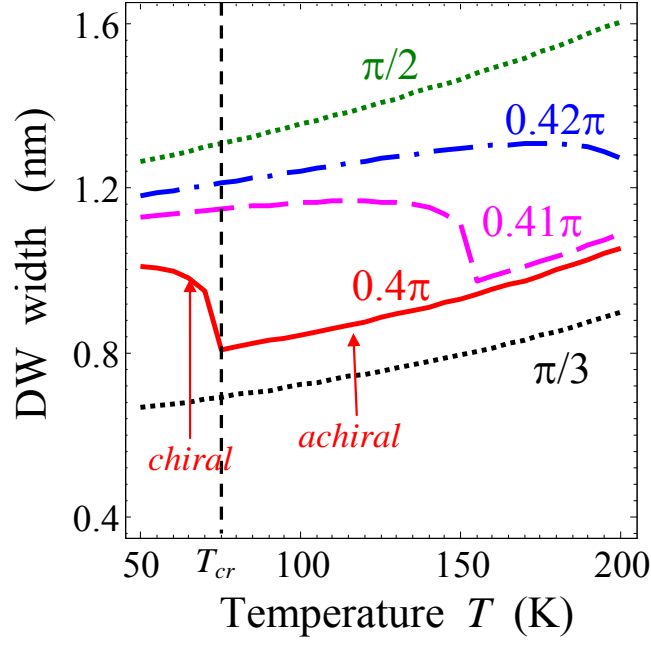

**Figure 5.** Temperature dependence of DW width calculated in rhombohedral ferroelectric phase of BTO for different rotation angles $\alpha = \pi/2, \pi/3, 0.4\pi, 0.41\pi, 0.42\pi$ specified near the curves.

## 3. IMPACT OF THE FLEXOELECTRIC COUPLING

To take into account the flexoelectric contribution we add the term

$$G_f = \frac{\tilde{F}_{ijkl}}{2}\left(\tilde{\sigma}_{ij}\frac{\partial \tilde{P}_k}{\partial \tilde{x}_l} - \tilde{P}_k \frac{\partial \tilde{\sigma}_{ij}}{\partial \tilde{x}_l}\right) \qquad (4)$$

into the Gibbs potential (1), where $\tilde{F}_{ijkl}$ is the flexoelectric tensor. Mathematically the flexoelectric coupling leads to the inhomogeneity in Euler-Lagrange equations: $-\tilde{F}_{12}(\partial\tilde{\sigma}_2/\partial\tilde{x}_1) - \tilde{F}_{13}(\partial\tilde{\sigma}_3/\partial\tilde{x}_1) - 2\tilde{F}_{14}(\partial\tilde{\sigma}_4/\partial\tilde{x}_1)$ in equation for $\tilde{P}_1$, $2\tilde{F}_{15}\partial\tilde{\sigma}_4/\partial\tilde{x}_1$ in equation for $\tilde{P}_2$, $\tilde{F}_{15}\partial\tilde{\sigma}_2/\partial\tilde{x}_1$ in equation for $\tilde{P}_3$ (see **Appendix S3**, **Suppl. Mat.**), which after all transformations in turn cause the following physical consequences.

Flexoelectric coupling introduces additional angular anisotropy for the DW structure and energy. One can see from **Fig. 6a** that the period of modulation of Neel component doubles, it is $2\pi/3$ for nonzero flexoelectric coupling and $\pi/3$ without flexoelectric coupling. Remarkable is that the "ground states" at $\alpha = m\pi/3$ stay equivalent, while the energy maxima at $\alpha = \pi/6 + m\pi/3$ for odd and even *m* become nonequivalent (**Fig. 6b**). This is seen from the different width of the area of chiral wall absolute stability and different height of the energy maximum. Note that the equivalence of the minima follows from the symmetry of the problem, which contains axis of third order along [111] and mirror plane {110}. For the maxima the situation is different since there is no mirror plane at {211} and the only symmetry operation is the axis of third order. That is why we observe two different triplets of maxima. Thus the flexoelectric coupling reveals the true symmetry of the problem, which was not reflected in the approximation without the coupling (see **Appendix S4, Suppl.Mat**). The flexoelectric contribution in the DW energy is comparable with energy anisotropy originated from electrostriction.

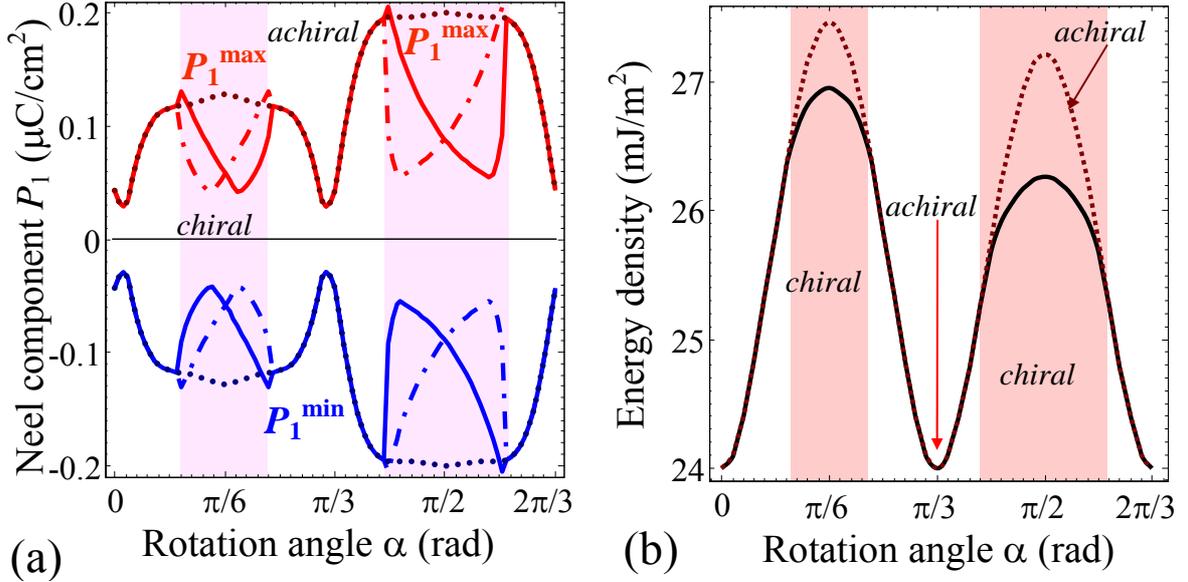

**Figure 6.** Angular dependence of maximal (**red** upper curves labeled $\widetilde{P}_i^{\max}$) and minimal (**blue** bottom curves labeled $\widetilde{P}_i^{\min}$) values of Neel (a) polarization component calculated in the vicinity of the DW in rhombohedral phase of BTO temperature 200 K, flexoelectric coefficients $F_{11}$= 2.46, $F_{12}$=0.48, $F_{44}$=0.05 in $10^{-11}C^{-1}m^3$ [52]; all designations are the same as in the Figure 2d. (b) Angular dependence of the 180-degree Ising-Bloch-Neel DW energy density calculated in rhombohedral phase of BTO; all designations are the same as in the Figure 3a.

Neel component of polarization appeared much more sensitive to the flexoelectric coupling than the Bloch one. Due to the coupling Neel component amplitude is nonzero for all rotation angles including $\alpha = m\pi/3$, while it is still minimal for this angle (compare **Figs. 2d** and **6a**). Thus flexoelectric coupling acts as additional and relatively strong source for the Neel polarization component.

## 4. C-AFM CONTRAST OF DW

Despite the origin of Neel polarization component presents a fundamental interest, its value is relatively small in comparison with Bloch and Ising components (compare the scale in **Fig. 2a,c** and **2b,d**), and the question about the experimental justification of our prediction naturally arises. Neel component leads to free carriers accumulation or depletion near the wall. Thus one possible way to verify theoretical results is to study local electronic properties of the wall by c-AFM [25-24].

### 4.1. Statement of the problem for the domain wall conductance

The conductivity enhancement in the domain wall is caused by the potential variation inside the wall. Keeping in mind realistic BTO with impurities we assume that the concentration of holes is negligible and the conductivity is purely of n-type [50]. The potential well/hump leads to higher/lower electron concentration in the DW due to the local band bending (see sketches in **Figs.7** for chiral and achiral walls).

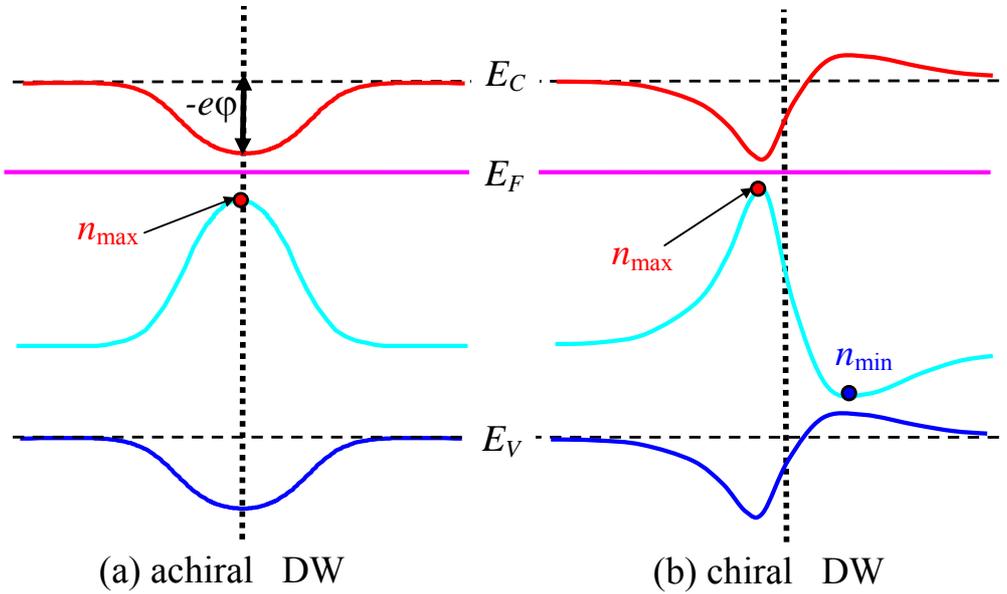

**Figure 7.** Sketches of local band bending for achiral (**a**) and chiral (**b**) walls, where the spatial regions with maximal ($n_{max}$) and minimal ($n_{min}$) electron density are indicated.

In general case one has to consider self-consistent problem and take into account potential redistribution due to the presence of electrons. However as we show a-posteriori by numerical simulations using the Poisson equation for electrostatic potential $\varphi$, for realistic charge carriers concentration in BTO the screening of the bound charge by electrons is negligible, thus their distribution can be found in "dielectric limit" with enough accuracy. In such approximation one can found the potential $\varphi$

$$\varphi(\tilde{x}_1) \approx \frac{1}{\varepsilon_0 \varepsilon_b} \int_{-\infty}^{\tilde{x}_1} d\tilde{x} \cdot \tilde{P}_1(\tilde{x}) \tag{5}$$

Free electron density $n(\tilde{x}_1)$ distribution can be estimated as [53]:

$$n(\tilde{x}_1) = \int_0^\infty d\varepsilon \cdot g_n(\varepsilon) f(\varepsilon + E_C - E_F - e\varphi(\tilde{x}_1)) \approx n_c \exp\left(\frac{E_F - E_C + e\varphi(\tilde{x}_1)}{k_B T}\right), \quad (6)$$

where $g_n(\varepsilon) = \sqrt{2m_n^3 \varepsilon}/(\pi^2 \hbar^3)$ is the energy density of states in the effective mass approximation, $m_n$ is the effective mass; $f(x) = (1 + \exp(x/k_B T))^{-1}$ is the Fermi-Dirac distribution function, $k_B = 1.3807 \times 10^{-23}$ J/K, $T$ is the absolute temperature, $E_F$ is Fermi level position, $E_C$ is the bottom of the conductive band, $e = 1.6 \times 10^{-19}$ C is the electron charge. Approximate equality in Eq.(6) corresponds to Boltzmann approximation for which the density of states in the conduction band $n_c = \frac{\sqrt{\pi m_n^3 k_B^3 T^3}}{\sqrt{2}\pi^2 \hbar^3}$. We checked that Boltzmann approximation works adequately here for $e|\varphi| \leq 5 k_B T$. Fermi level position $E_F(T)$ in the frame of our approximation may be found in terms of electron concentration in the single-domain region $n_0(T) = n_c \exp\left(\frac{E_F - E_C}{k_B T}\right)$ as

$$E_F(T) = E_C + k_B T \ln(n_0/n_c)$$

In our model we do not take into account deformation potential [54, 55], because in the model case of the non-degenerated simple band structure and within the validity of effective mass approximation, the shallow donor level and the conductive band edge are shifted as a whole with the strain [56].

Another assumption that we make is the continuity of the band structure across the DW. Rigorously speaking the potential barrier or well, $\varphi(\tilde{x}_1)$, should be included into the quantum-mechanical treatment since quantization should exist in the direction transverse to the wall, which has thickness ~ 1 nm. Here we are interested in conductivity along the DW where no quantization occurs. We calculate the potential jump $\varphi(\tilde{x}_1)$ within continuum media theory and hope that results obtained for the carrier's accumulation/depletion across the DW are qualitatively valid and will be justified by rigorous quantum-mechanical approach elsewhere.

Results of the numerical modeling for the DW polarization vector structure, electric potential and charge carriers redistribution across the domain wall are discussed below.

### 4.2. Phase transition detection in DW by c-AFM contrast

Since Neel component profile is anti-symmetric for achiral DW, corresponding potential barrier $\varphi(\tilde{x}_1)$ is symmetric, while it can be asymmetric for achiral DWs. Symmetric barriers $\varphi(\tilde{x}_1)$ accumulate electrons with maximal density $n_{max}$ (**Fig.7a**). Asymmetric double barriers can attract the electrons in some spatial regions with maximal density $n_{max}$ and repulse them from the other regions with minimal density $n_{min}$ (**Fig.7b**). The most intriguing situation can appear in the point of the wall chiral-achiral phase transition, i.e. at rotation angles around the critical ones, $\alpha_{cr}$. The chiral-achiral phase transition can be revealed by local c-AFM measurements of the cylindrical walls, since c-AFM contrast is regarded proportional to the relative electron density $n(\widetilde{x}_1)/n_0$ [28]. **Figures 8** illustrate the rotation anisotropy of the relative density $n(\widetilde{x}_1)/n_0$. Exactly two sharp maxima on $n_{max}$ and breaks $n_{min}$ on **Fig. 8** corresponds to the chiral-achiral phase transitions occurred at $\alpha_{cr}^m \approx \pi/6 \pm \pi/12 + m\pi/3$. Without flexoelectric coupling c-AFM contrast is equal to unity for the angles $\alpha = m\pi/3$ corresponding to the absence of Neel component (see **Fig. 8a and 2d**). Flexoelectric coupling leads to nonzero Neel component for all $\alpha$ and thus to nonzero contrast; also it slightly shifts the critical angles and create the symmetric potential structure well-barrier-well around rotation angle $\pi/3$ (see **Fig. 8b** and **6a**). Results shown in the **Figure 8** for rhombohedral BTO look principally different from the ones presented in Ref. [17] for rhombohedral BiFeO$_3$. This difference may be explained because in BiFeO$_3$ the domain walls are only of achiral type, and the coupling between Neel and Bloch components is not so strong.

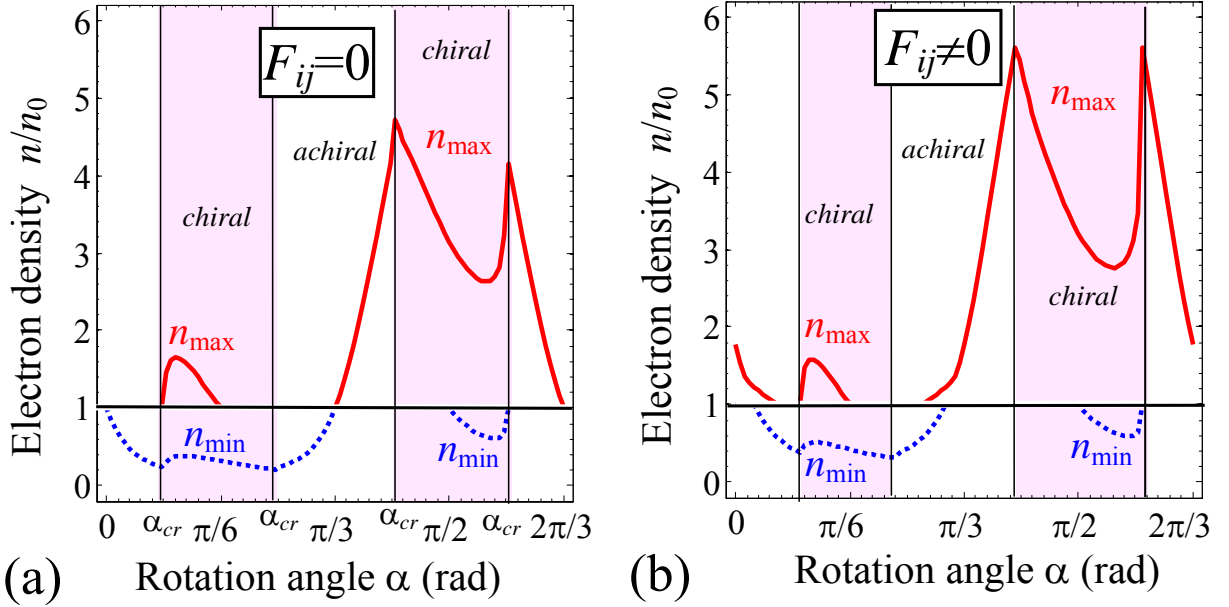

**Figure 8.** Relative maximal $n_{max}/n_0$ and minimal $n_{min}/n_0$ electron density vs. the DW rotation angle $\alpha$ calculated in BTO at 200 K without flexoelectric coupling $F_{ij} = 0$ (a) and with flexoelectric coupling $F_{11}= 2.46$, $F_{12}=0.48$, $F_{44}=0.05$ in $10^{-11}$C$^{-1}$m$^3$ (b). Temperature 200 K.

The potential barrier (or well) $\varphi(\widetilde{x}_1)$ and electron density $n(\widetilde{x}_1)$ profiles calculated for different wall orientations are shown in **Fig. 9**. The walls oriented near $\alpha_{cr} \approx 5\pi/12$ corresponding to chiral-achiral phase transition have maximal electron accumulation, because Neel polarization component is maximal there (see **Fig 6a**). One can see from the **Figure 9b** that maximal electron density $n(\tilde{x}_1)$ is about 4 times higher than the electron density $n_0$ in the single domain region of BTO. This means that the wall relative conductivity at the wall becomes at least several times higher than in the single-domain region, i.e. the ratio $\sigma_{max}/\sigma(\pm\infty) \gg 1$. Such contrast is pronounced and thus can be easily detected by c-AFM.

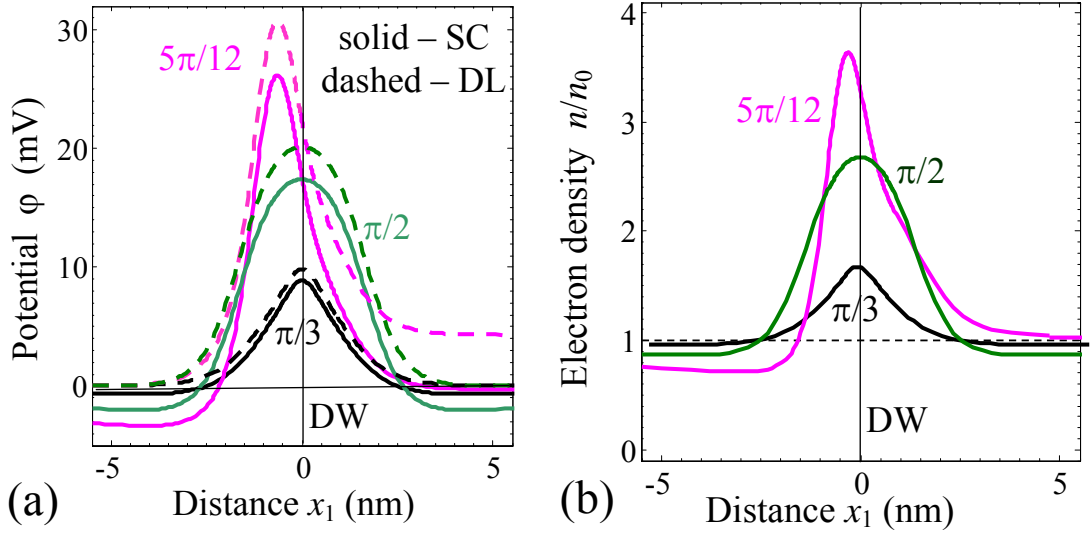

**Figure 9.** Profiles of potential $\varphi(\tilde{x}_1)$ (a), and relative electron density $n(\tilde{x}_1)/n_0$ (b) calculated across the DW for rotation angles $\alpha = \pi/3, 5\pi/12, \pi/2$ (specified near the curves), temperature 200 K, flexoelectric coefficients $F_{11}= 2.46$, $F_{12}=0.48$, $F_{44}=0.05$ in $10^{-11}$C$^{-1}$m$^3$ and BTO parameters listed in the **Table S1**. Solid curves in plot (a) correspond to full-scale calculations with account of semiconducting properties: $n_0 = 3\times10^{22}$m$^{-3}$ (SC); dashed curves are calculated in dielectric limit $n_0 = 0$ (DL).

Temperature dependence of the c-AFM contrast of chiral walls, calculated as relative carrier density, $n_{max}(\tilde{x}_1)/n_0$, is shown in **Fig. 10a.** The contrast increases with the temperature decrease exponentially in the Boltzmann approximation, where $n \approx n_0(T)\exp(e\varphi/k_BT)$. At temperatures lower than 50 K the wall c-AFM contrast between the wall and the single-domain region becomes more 10 times even for the case of the most weakly conducting walls corresponding to rotation angles $\alpha = \pi/3 + 2m\pi/3$. The angle $\alpha = \pi/3$ corresponds to the DW with minimal energy. For other rotation angles (e.g. for $\alpha = 0.39\pi, \pi/2$) the c-AFM contrast can be 50-500 times higher than the single-domain one. The angle $\alpha = \pi/2$ corresponds to the DW with maximal energy. Note that the concentration $n_0$ strongly decreases with temperature decrease as shown in **Fig. S4, Suppl. Mat.**

We suggest that the phase transition in the wall structure can be detected by the jump on the c-AFM contrast temperature dependence. Such jump takes place for example at $T^*_{cr} \sim 105$ K for the angle $\alpha^*_{cr} = 0.39\pi$, which exactly corresponds to the abrupt phase transition in the wall structure, which is slightly shifted from the value $\alpha_{cr} = 5\pi/12$ by the flexoelectric effect. Strong correlation between the c-AFM contrast (**Figs.10a**), maximal potential at the wall $\varphi_{max}(\tilde{x}_1)$ (**Figs.10b**), Bloch (**Figs.10c**) and Neel (**Figs.10d**) components can be predicted from our study. Thus we hope that our calculations can stimulate further experimental c-AFM studies of the wall conduction in BTO, other ferroelectrics and multiferroics in a wide temperature range, since the studies can give insight to the wall polar structure and conductivity correlations, as well as quantitative information of the flexoelectric coupling strength.

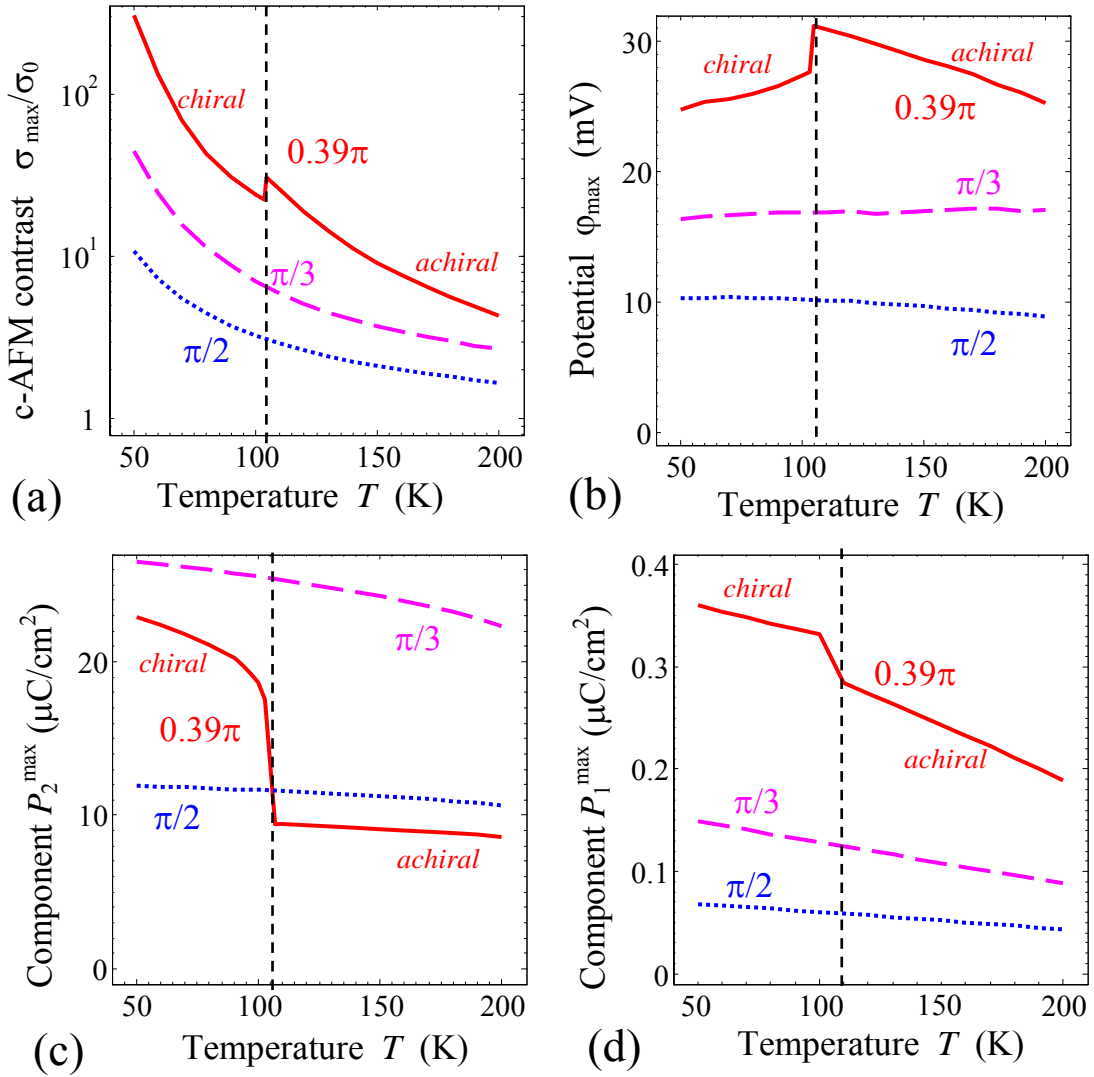

**Figure 10.** (a) Temperature dependence of the maximal c-AFM contrast $\sigma_{max}/\sigma_0$ (a); potential jump $\varphi_{max}$ (b); Bloch (c) and Neel (d) components at DW in rhombohedral ferroelectric phase of BTO calculated for different rotation angles $\alpha = \pi/2, \pi/3, 0.39\pi$ specified near the curves. Note, that the angles $\alpha = \pi/3$ and $\alpha = \pi/2$ correspond to the DW with minimal and maximal energy, and one can see that even in the limiting cases the DWs are more conductive than the bulk.

## 5. SUMMARY

In the frame of LGD theory domain wall structure and energy are investigated in rhombohedral BTO as functions of 180-degree domain wall orientation. It is shown that there are six energetically favorable wall orientations corresponding to {110}-planes. {211} orientations correspond to energy maxima. The minima are equivalent, while maxima are equivalent only disregarding the flexoelectric effect and split into two triplets under its presence. This may be explained by the presence of mirror plane at {110} and its absence at {211}. Another impact of the flexoeffect is that Neel component is nonzero for any wall orientation (0 for {110}-wall in its absence). Thus flexoelectric effect reveals the true symmetry of the problem.

Domain walls are shown to be of mixed Ising-Bloch-Neel type for all orientations. However domain walls with {211} and {110} orientations are shown to have sufficiently different structures, achiral and chiral. The phase transition from achiral to chiral state can be achieved varying wall's tilt angle and for some orientations by temperature change at constant orientation

We suggest detecting the structural phase change inside domain walls by c-AFM contrast due to the correlation of the domain wall structure and free charge accumulation, driven by depolarizing field. Depending on the temperature and orientation the conductivity of the wall may be one or even two orders of magnitude higher than in the single-domain region. Achiral-chiral phase transition in the wall is accompanied with rapid change of the wall c-AFM contrast. In this context c-AFM appears to be promising tool for the detection of structural phase transitions inside domain walls.

**Acknowledgements**

A.N.M. and E.A.E. gratefully acknowledge multiple discussions, useful suggestions and critical remarks from Prof. N.V. Morozovsky. E.A.E. and A.N.M. are thankful to NAS Ukraine for support. P.V.Y., A.K.T and N.S. acknowledge Swiss National Science foundation for financial support. S.V.K. research is supported by the US Department of Energy, Basic Energy Sciences, Materials Sciences and Engineering Division.

# SUPPLEMENTAL MATERIALS TO
# CORRELATION BETWEEN STRUCTURE AND C-AFM CONTRAST OF 180-DEGREE DOMAIN WALLS IN RHOMBOHEDRAL BaTi0$_3$

**Table S1.** Material parameters for bulk ferroelectric BaTiO$_3$

| coefficient | BaTiO$_3$ (collected and recalculated mainly from Ref. [a, b]) |
|---|---|
| **Symmetry** | tetragonal |
| $\varepsilon_b$ | 7 (Ref. [b]) |
| $a_i$ (C$^{-2}$·mJ) | $a_1$=3.34($T$–381)×10$^5$ (at 293°K –2.94×10$^7$) |
| $a_{ij}$ (C$^{-4}$·m$^5$J) | $a_{11}$= 4.69($T$–393)×10$^6$–2.02×10$^8$, $a_{12}$= 3.230×10$^8$, (at 293°K $a_{11}$= –6.71×10$^8$ $a_{12}$= 3.23×10$^8$) |
| $a_{ijk}$ (C$^{-6}$·m$^9$J) | (at 293°K $a_{111}$= 82.8×10$^8$, $a_{112}$=44.7×10$^8$, $a_{123}$=49.1×10$^8$) $a_{111}$= –5.52($T$–393)×10$^7$+2.76×10$^9$ $a_{112}$=4.47×10$^9$ $a_{123}$=4.91×10$^9$ |
| $Q_{ij}$ (C$^{-2}$·m$^4$) | $Q_{11}$=0.11, $Q_{12}$= –0.043, $Q_{44}$=0.059 |
| $s_{ij}$ (×10$^{-12}$ Pa$^{-1}$) | $s_{11}$=8.3, $s_{12}$= –2.7, $s_{44}$=9.24 |
| $g_{ij}$ (×10$^{-10}$C$^{-2}$m$^3$J) | $g_{11}$=5.1, $g_{12}$= –0.2, $g_{44}$= 0.2 [c] |
| $F_{ij}$ (×10$^{-11}$C$^{-1}$m$^3$) | ~100 (estimated from measurements of Ref. [d]) $F_{11}$= +2.46, $F_{12}$=0.48, $F_{44}$=0.05 (recalculated from [e] using $F_{\alpha\gamma}=f_{\alpha\beta}s_{\beta\gamma}$) |

[a] A.J. Bell. J. Appl. Phys. 89, 3907 (2001).

[b] J. Hlinka and P. Márton, Phys. Rev. B 74, 104104 (2006).

[c] P. Marton, I. Rychetsky, and J. Hlinka. Phys. Rev. B 81, 144125 (2010).

[d] W. Ma and L. E. Cross, Appl. Phys. Lett., 88, 232902 (2006).

[e] I. Ponomareva, A. K. Tagantsev, L. Bellaiche. Phys.Rev B 85, 104101 (2012).

### Appendix S1. Euler-Lagrange equations for polarization components
#### *S.1a. 180-degree walls in rhombohedral ferroelectric phase*

Coordinates transformation for 180-degree uncharged domain walls in rhombohedral phase is

$$\widetilde{x}_1 = \sqrt{\frac{2}{3}} \sin\alpha\, x_1 + x_2\left(\frac{\cos\alpha}{\sqrt{2}} - \frac{\sin\alpha}{\sqrt{6}}\right) - x_3\left(\frac{\cos\alpha}{\sqrt{2}} + \frac{\sin\alpha}{\sqrt{6}}\right), \quad (S.1a)$$

$$\widetilde{x}_2 = -\sqrt{\frac{2}{3}} \cos\alpha\, x_1 + x_2\left(\frac{\cos\alpha}{\sqrt{6}} + \frac{\sin\alpha}{\sqrt{2}}\right) + x_3\left(\frac{\cos\alpha}{\sqrt{6}} - \frac{\sin\alpha}{\sqrt{2}}\right), \quad (S.1b)$$

$$\widetilde{x}_3 = \frac{x_1 + x_2 + x_3}{\sqrt{3}}, \quad (S.1c)$$

where $\alpha$ is the rotation angle with respect to the cube spatial diagonal.

Without flexoelectric coupling Euler-Lagrange equations for polarization components depending only on $\tilde{x}_1$ have the form:

$$2a_1\tilde{P}_1 + 4\tilde{a}_{11}\tilde{P}_1^3 + 2\tilde{a}_{12}\tilde{P}_2^2\tilde{P}_1 + 2\tilde{a}_{13}\tilde{P}_3^2\tilde{P}_1 + 3\tilde{a}_{15}(\tilde{P}_1^2 - \tilde{P}_2^2)\tilde{P}_3 - 6\tilde{a}_{24}\tilde{P}_1\tilde{P}_2\tilde{P}_3 - \tilde{g}_{11}\frac{\partial^2 \tilde{P}_1}{\partial \tilde{x}_1^2}$$
$$-\tilde{g}_{15}\frac{\partial^2 \tilde{P}_3}{\partial \tilde{x}_1^2} - 2(\tilde{Q}_{12}\tilde{\sigma}_2 + \tilde{Q}_{13}\tilde{\sigma}_3 + \tilde{Q}_{14}\tilde{\sigma}_4)\tilde{P}_1 + 2\tilde{Q}_{15}\tilde{\sigma}_4\tilde{P}_2 + \tilde{Q}_{15}\tilde{\sigma}_2\tilde{P}_3 = \tilde{E}_1^d$$
(S.2a)

$$2a_1\tilde{P}_2 + 4\tilde{a}_{11}\tilde{P}_2^3 + 2\tilde{a}_{12}\tilde{P}_2^2\tilde{P}_1 + 2\tilde{a}_{13}\tilde{P}_3^2\tilde{P}_1 + 3\tilde{a}_{24}(-\tilde{P}_1^2 + \tilde{P}_2^2)\tilde{P}_3 - 6\tilde{a}_{15}\tilde{P}_1\tilde{P}_2\tilde{P}_3 - \tilde{g}_{66}\frac{\partial^2 \tilde{P}_2}{\partial \tilde{x}_1^2}$$
$$-\tilde{g}_{14}\frac{\partial^2 \tilde{P}_3}{\partial \tilde{x}_1^2} - 2(\tilde{Q}_{11}\tilde{\sigma}_2 + \tilde{Q}_{13}\tilde{\sigma}_3 - \tilde{Q}_{14}\tilde{\sigma}_4)\tilde{P}_2 + 2\tilde{Q}_{15}\tilde{\sigma}_4\tilde{P}_1 + (\tilde{Q}_{14}\tilde{\sigma}_2 - \tilde{Q}_{44}\tilde{\sigma}_4)\tilde{P}_3 = 0$$
(S.2b)

$$2a_1\tilde{P}_3 + 4\tilde{a}_{33}\tilde{P}_3^3 + 2\tilde{a}_{13}(\tilde{P}_1^2 + \tilde{P}_2^2)\tilde{P}_3 + \tilde{a}_{15}(\tilde{P}_1^2 - 3\tilde{P}_2^2)\tilde{P}_1 + \tilde{a}_{24}(-3\tilde{P}_1^2 + \tilde{P}_2^2)\tilde{P}_2 - \tilde{g}_{44}\frac{\partial^2 \tilde{P}_3}{\partial \tilde{x}_1^2}$$
$$-\tilde{g}_{14}\frac{\partial^2 \tilde{P}_2}{\partial \tilde{x}_1^2} - \tilde{g}_{15}\frac{\partial^2 \tilde{P}_1}{\partial \tilde{x}_1^2} - 2(\tilde{Q}_{13}\tilde{\sigma}_2 + \tilde{Q}_{33}\tilde{\sigma}_3)\tilde{P}_3 + (\tilde{Q}_{14}\tilde{\sigma}_2 - \tilde{Q}_{44}\tilde{\sigma}_4)\tilde{P}_2 + \tilde{Q}_{15}\tilde{\sigma}_2\tilde{P}_1 = 0$$
(S.2c)

Coefficients $a_1$, $a_{12}$ and $\tilde{a}_{11} = (2a_{11} + a_{12})/4$, $\tilde{a}_{12} = (2a_{11} + a_{12})/2$, $\tilde{a}_{13} = 2a_{11}$, $\tilde{a}_{33} = (a_{11} + a_{12})/3$, $\tilde{a}_{15} = -\sqrt{2}\sin(3\alpha)(2a_{11} - a_{12})/3$, $\tilde{a}_{24} = -\sqrt{2}\cos(3\alpha)(2a_{11} - a_{12})/3$ are the LGD-potential expansion coefficients. The gradient coefficients are and $\tilde{g}_{11} = \frac{g_{11} + g_{12} + 2g_{44}}{2}$, $\tilde{g}_{44} = \frac{g_{11} - g_{12} + g_{44}}{3}$, $\tilde{g}_{66} = \frac{g_{11} - g_{12} + 4g_{44}}{6}$, $\tilde{g}_{14} = \cos(3\alpha)\frac{g_{11} - g_{12} - 2g_{44}}{3\sqrt{2}}$, $\tilde{g}_{15} = -\sin(3\alpha)\frac{g_{11} - g_{12} - 2g_{44}}{3\sqrt{2}}$. Rotated tensors components are listed in the **Table S2**.

**Table S2.** Dependence of the tensors and other coefficients on the wall rotation angle α of 180-degree domain wall in the **rhombohedral phase** (taken from Ref.[**Ошибка! Закладка не определена.**]).

| Elastic compliances $\tilde{s}_{ij}$ in rotated frame $\{\tilde{x}_1, \tilde{x}_2, \tilde{x}_3\}$ | $\tilde{s}_{11} = \frac{2(s_{11}+s_{12})+s_{44}}{4}$, $\tilde{s}_{12} = \frac{2s_{11}+10s_{12}-s_{44}}{12}$, $\tilde{s}_{13} = \frac{2s_{11}+4s_{12}-s_{44}}{6}$, $\tilde{s}_{33} = \frac{s_{11}+2s_{12}+s_{44}}{3}$, $\tilde{s}_{66} = \frac{2(s_{11}-s_{12})+2s_{44}}{3}$, $\tilde{s}_{44} = \frac{4(s_{11}-s_{12})+s_{44}}{3}$, $\tilde{s}_{14} = \frac{\cos(3\alpha)}{3\sqrt{2}}(2(s_{11}-s_{12})-s_{44})$, $\tilde{s}_{15} = -\frac{\sin(3\alpha)}{3\sqrt{2}}(2(s_{11}-s_{12})-s_{44})$ |
|---|---|
| Electro-striction tensor components in | $\tilde{Q}_{11} = \frac{2(Q_{11}+Q_{12})+Q_{44}}{4}$, $\tilde{Q}_{12} = \frac{2Q_{11}+10Q_{12}-Q_{44}}{12}$, $\tilde{Q}_{13} = \frac{2Q_{11}+4Q_{12}-Q_{44}}{6}$, $\tilde{Q}_{33} = \frac{Q_{11}+2Q_{12}+Q_{44}}{3}$, $\tilde{Q}_{66} = \frac{2(Q_{11}-Q_{12})+2Q_{44}}{3}$, $\tilde{Q}_{44} = \frac{4(Q_{11}-Q_{12})+Q_{44}}{3}$, |

| $\{\tilde{x}_1, \tilde{x}_2, \tilde{x}_3\}$ | $\tilde{Q}_{14} = \dfrac{\cos(3\alpha)}{3\sqrt{2}}(2(Q_{11}-Q_{12})-Q_{44})$, $\tilde{Q}_{15} = -\dfrac{\sin(3\alpha)}{3\sqrt{2}}(2(Q_{11}-Q_{12})-Q_{44})$ |

## Appendix S2. Solution for elastic stresses in the rhombohedral phase

In the case of $\tilde{x}_1$-dependent solution, compatibility relation $e_{ikl}e_{jmn}\left(\partial^2 \tilde{u}_{ln}/\partial \tilde{x}_k \partial \tilde{x}_m\right) = 0$ leads to the conditions of constant strains $\tilde{u}_2 = const$, $\tilde{u}_3 = const$, $\tilde{u}_4 = const$, while general form dependences like $\tilde{u}_1 = \tilde{u}_1(\tilde{x}_1)$, $\tilde{u}_5 = \tilde{u}_5(\tilde{x}_1)$ and $\tilde{u}_6 = \tilde{u}_6(\tilde{x}_1)$ do not contradict to these relations. Mechanical equilibrium conditions $\partial \tilde{\sigma}_{ij}/\partial \tilde{x}_i = 0$ could be written as $\partial \tilde{\sigma}_1/\partial \tilde{x}_1 = 0$, $\partial \tilde{\sigma}_5/\partial \tilde{x}_1 = 0$, $\partial \tilde{\sigma}_6/\partial \tilde{x}_1 = 0$. Since $\tilde{\sigma}_{ij}(\tilde{x}_1 \to \pm\infty) = 0$, one obtains $\tilde{\sigma}_1 = \tilde{\sigma}_5 = \tilde{\sigma}_6 = 0$. Elastic strains can be found from equations of state $\partial \tilde{G}/\partial \tilde{\sigma}_{ij} = -\tilde{u}_{ij}$. After substitution of corresponding elastic stresses acquire the form:

$$\tilde{\sigma}_2 = \frac{\tilde{s}_{44}(\tilde{s}_{33}U_2 - \tilde{s}_{13}U_3) + \tilde{s}_{14}\tilde{s}_{33}U_4}{(\tilde{s}_{11}\tilde{s}_{33} - \tilde{s}_{13}^2)\tilde{s}_{44} - \tilde{s}_{14}^2\tilde{s}_{33}}, \quad \tilde{\sigma}_3 = \frac{-\tilde{s}_{44}\tilde{s}_{13}U_2 + (\tilde{s}_{44}\tilde{s}_{11} - \tilde{s}_{14}^2)U_3 - \tilde{s}_{14}\tilde{s}_{13}U_4}{(\tilde{s}_{11}\tilde{s}_{33} - \tilde{s}_{13}^2)\tilde{s}_{44} - \tilde{s}_{14}^2\tilde{s}_{33}}, \qquad \text{(S.4a)}$$

$$\tilde{\sigma}_4 = \frac{\tilde{s}_{14}(\tilde{s}_{33}U_2 - \tilde{s}_{13}U_3) + (\tilde{s}_{11}\tilde{s}_{33} - \tilde{s}_{13}^2)U_4}{(\tilde{s}_{11}\tilde{s}_{33} - \tilde{s}_{13}^2)\tilde{s}_{44} - \tilde{s}_{14}^2\tilde{s}_{33}}, \quad \tilde{\sigma}_1 = \tilde{\sigma}_5 = \tilde{\sigma}_6 = 0, \qquad \text{(S.4b)}$$

Where the functions $U_3$ and $U_2$ are introduced as

$$U_2 = -\tilde{Q}_{12}\tilde{P}_1^2 - \tilde{Q}_{11}\tilde{P}_2^2 + \tilde{Q}_{13}\left((\tilde{P}_3^S)^2 - \tilde{P}_3^2\right) + \tilde{Q}_{14}\tilde{P}_2\tilde{P}_3 + \tilde{Q}_{15}\tilde{P}_1\tilde{P}_3, \qquad \text{(S.5a)}$$

$$U_3 = -\tilde{Q}_{13}\left(\tilde{P}_1^2 + \tilde{P}_2^2\right) + \tilde{Q}_{33}\left((\tilde{P}_3^S)^2 - \tilde{P}_3^2\right), \qquad \text{(S.5b)}$$

and

$$U_4 = -\tilde{Q}_{44}\tilde{P}_2\tilde{P}_3 - \tilde{Q}_{14}\left(\tilde{P}_1^2 - \tilde{P}_2^2\right) + 2\tilde{Q}_{15}\tilde{P}_1\tilde{P}_2 \qquad \text{(S.5c)}$$

## Appendix S3. Flexoelectric coupling impact on the Euler-Lagrange equations and elastic stresses

1) Flexoelectric coupling leads to the additional terms in Euler-Lagrange equations: $-\tilde{F}_{12}(\partial \tilde{\sigma}_2/\partial \tilde{x}_1) - \tilde{F}_{13}(\partial \tilde{\sigma}_3/\partial \tilde{x}_1) - 2\tilde{F}_{14}(\partial \tilde{\sigma}_4/\partial \tilde{x}_1)$ in the right-hand-side of equation (S.1a) for $\tilde{P}_1$, $2\tilde{F}_{15}\partial \tilde{\sigma}_4/\partial \tilde{x}_1$ in the right-hand-side of equation (S.1b) for $\tilde{P}_2$, $\tilde{F}_{15}\partial \tilde{\sigma}_2/\partial \tilde{x}_1$ in the right-hand-side of equation (S.1b) for $U_4$.

2) Flexoelectric coupling leads to the additional term $\tilde{F}_{12}\dfrac{\partial \tilde{P}_1}{\partial \tilde{x}_1} - \tilde{F}_{15}\dfrac{\partial \tilde{P}_3}{\partial \tilde{x}_1}$ in the equation (S.5a) for $U_2$, $+\tilde{F}_{13}\dfrac{\partial \tilde{P}_3}{\partial \tilde{x}_1}$ in the equation (S.5b) for $U_3$, $2\tilde{F}_{14}\dfrac{\partial \tilde{P}_1}{\partial \tilde{x}_1} - 2\tilde{F}_{15}\dfrac{\partial \tilde{P}_2}{\partial \tilde{x}_1}$ in the right-hand-side of equation (S.5c) for $U_4$.

3) Flexoelectric tensor components in rotated frame $\{\tilde{x}_1, \tilde{x}_2, \tilde{x}_3\}$ are $\tilde{F}_{11} = \dfrac{F_{11} + F_{12} + F_{44}}{2}$, $\tilde{F}_{12} = \dfrac{F_{11} + 5F_{12} - F_{44}}{6}$, $\tilde{F}_{13} = \dfrac{F_{11} + 2F_{12} - F_{44}}{3}$, $\tilde{F}_{33} = \dfrac{F_{11} + 2(F_{12} + F_{44})}{3}$, $\tilde{F}_{66} = \dfrac{F_{11} - F_{12} + 2F_{44}}{3}$, $\tilde{F}_{44} = \dfrac{2(F_{11} - F_{12}) + F_{44}}{3}$, $\tilde{F}_{14} = \dfrac{\cos(3\alpha)}{3\sqrt{2}}(F_{11} - F_{12} - F_{44})$, $\tilde{F}_{15} = -\dfrac{\sin(3\alpha)}{3\sqrt{2}}(F_{11} - F_{12} - F_{44})$.

### Appendix S4. Flexoelectric coupling reveals the wall symmetry

Two types of purely Ising 180° DW in 3m ferroelectric are shown in Fig. S1. The domain walls (DW) in rhombohedral perovskites, viewed along {111}, with two orientations different on 180° are shown in Fig. S1. One structure could be transformed into another by appropriate translation along the domain wall. The transformation should be described by the spatial group theory, rather than point group one. For "up" and "down" polarization the structure (and hence the energy) of the walls rotated on 30° and 90° should be the same.

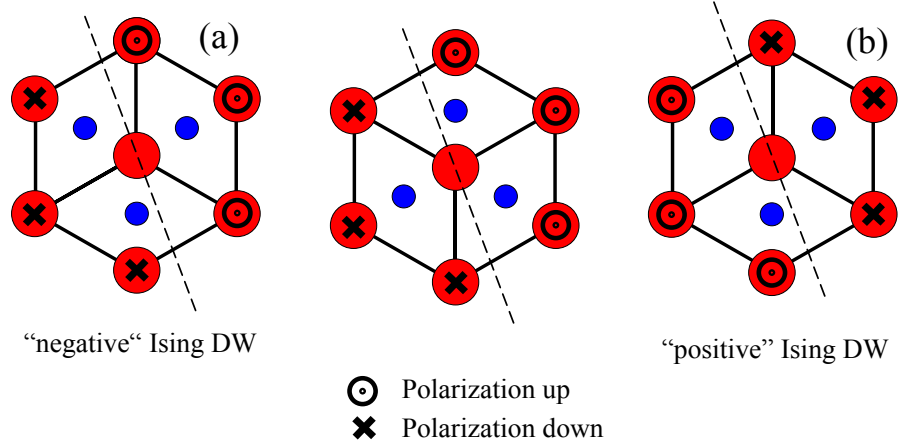

"negative" Ising DW                         "positive" Ising DW

⊙ Polarization up
✖ Polarization down

**Figure S1.** Two types of purely Ising 180° DW in 3m ferroelectric.

Flexoelectric coupling leads to the appearance of the small mixed parity Neel component perpendicular to the wall that weakly charges the wall and transforms from the head-to-head to tail-to-tail configuration under the wall rotation on 180°. Flexoelectric effect acts as an external field in the right-hand-side of the Euler-Lagrange equations. But it is not the reason for the energy difference for 30° and 90°-rotated walls. The quantitative reason is that under such transformation some of the flexoelectric tensor components ($F_{1322}$, $F_{1311}$, $F_{1312}$, $F_{2322}$,) change their signs with wall rotation from 30° to 90°. Since one of them ($F_{1322}$) couples gradient of $P_3$ ("Ising component") with quadratic function of all the polarization components, it gives contribution to energy with different signs for these two orientations (despite the amplitude of Neel component also changes under such transformation the linear contribution of $P_1$ gradient into the DW energy has much weaker orientation dependence).

One of the consequences of such phenomena is that one has to distinguish between DW with different signs of gradient of Ising polarization component (i.e. "negative" DW with $dP_3/dx_1 <0$ and "positive" with $dP_3/dx_1 >0$) in rhombohedral (3m) ferroelectrics. Note that in classical DW model that deals with 4mm point group (or with one component polarization at least) there is symmetry element (180° rotation around polar axis) that transforms "negative" DW into "positive" DW and hence their energies are the same (and there are no reasons to distinguish them). However, the presence of flexoelectric coupling is essential here, since it seems to be the only way to take into account the terms linear in $dP_3/dx_1$ in the free energy. The difference in energy between positive and negative DW is about 0.7 mJ/m$^2$ for rotation angle $\pi/6$. It seems that the difference will be absent for the angles multiple to $\pi/3$ (see **Figure S2).**

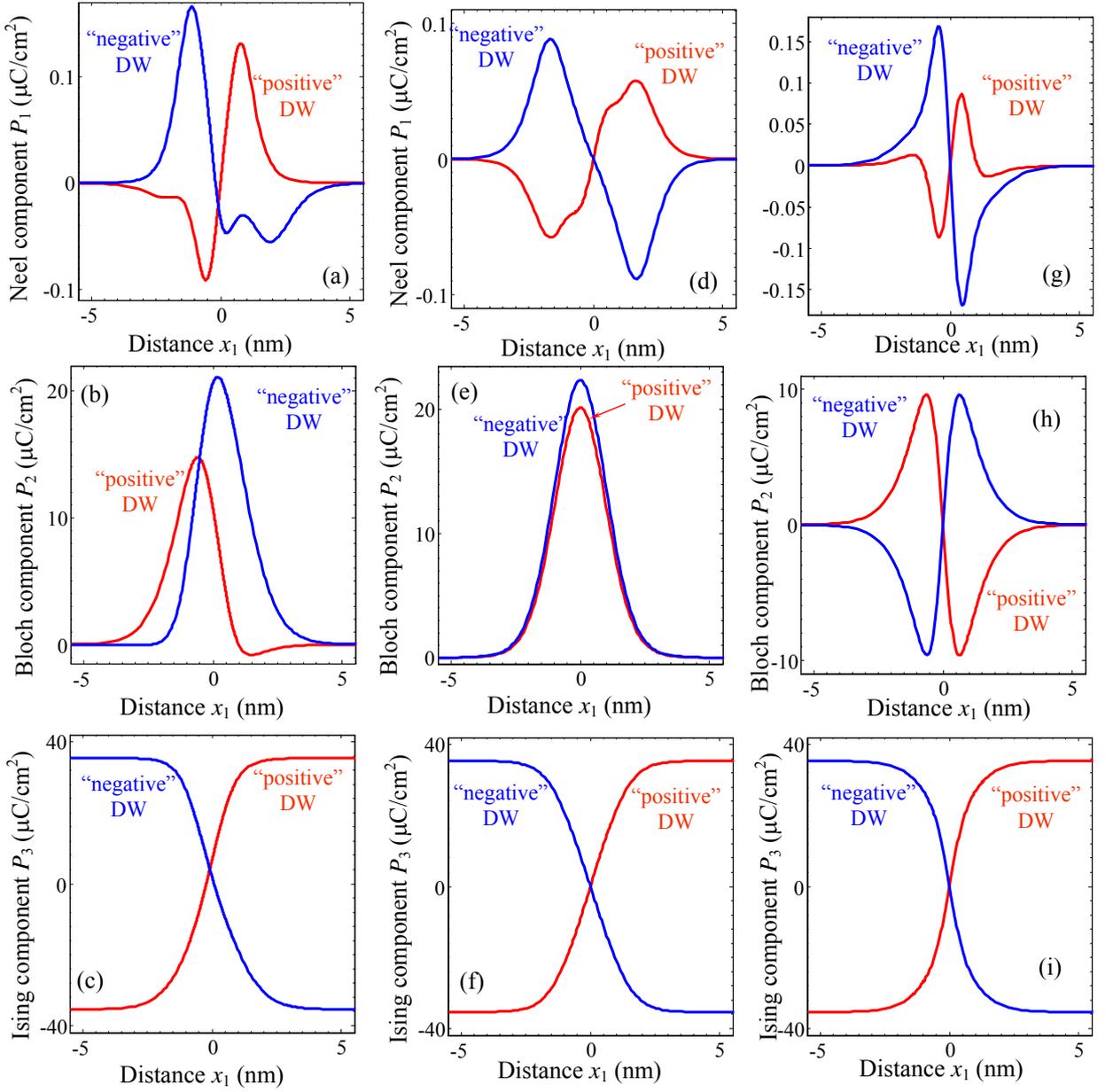

**Figure S2.** Material: BTO, temperature 200 K. Angles: π/10 (a, b, c); π/6 (d, e, f), π/24 (g, h, i), Flexoelectric coupling is included.

## Appendix S5. Electronic properties of DW

Considering *n*-type BaTiO$_3$, one should find the depolarization field $\widetilde{E}_1 = -\partial\varphi/\partial\widetilde{x}_1$ from the Poisson equation for electrostatic potential φ

$$\varepsilon_0\varepsilon_b \frac{\partial^2\varphi}{\partial\widetilde{x}_1^2} = \frac{\partial\widetilde{P}_1}{\partial\widetilde{x}_1} - e\left(N_d^+ - n\right) \quad \text{(S.6)}$$

Where $e=1.6\times10^{-19}$ C the electron charge, *n* is the concentration of the electrons in the conduction band; $N_d^+$ is the concentration of ionized donors. Hereinafter acceptors are regarded absent and holes concentration is regarded negligibly small in comparison with the concentration of electrons, which are improper carriers in *n*-type BaTiO$_3$. Free electron density $n(\tilde{x}_1)$ distribution can be estimated as:

$$n(\tilde{x}_1) = \int_0^\infty d\varepsilon \cdot g_n(\varepsilon) f(\varepsilon + E_C - E_F - e\varphi(\tilde{x}_1)) \approx n_0(T) \exp\left(\frac{e\varphi(\tilde{x}_1)}{k_B T}\right), \quad \text{(S.7a)}$$

where $g_n(\varepsilon) = \sqrt{2m_n^3 \varepsilon}/(\pi^2 \hbar^3)$ is density of states in the effective mass approximation, $m_n$ is the effective mass; $f(x) = (1 + \exp(x/k_B T))^{-1}$ is the Fermi-Dirac distribution function, $k_B = 1.3807 \times 10^{-23}$ J/K, $T$ is the absolute temperature, Fermi level position is $E_F$, the bottom of the conductive band is $E_C$. Approximate equality in Eq.(6a) corresponds to Boltzmann approximation for which $n_0(T) = \frac{\sqrt{\pi m_n^3 k_B^3 T^3}}{\sqrt{2}\pi^2 \hbar^3} \exp\left(\frac{E_F - E_C}{k_B T}\right)$. We checked that Boltzmann approximation works adequately here for $e|\varphi| \leq 5 k_B T$.

Concentration of almost immobile ionized donors in the Boltzmann approximation is

$$N_d^+(\tilde{x}_1) = N_{d0}^+(T) \exp\left(-\frac{e\varphi(\tilde{x}_1)}{k_B T}\right). \quad \text{(S.7b)}$$

Concentration of ionized donor centers far from the DW is $N_{d0}^+(T) = N_d^0 f(E_F - E_d) \approx N_d^0 \exp\left(\frac{E_d - E_F}{k_B T}\right)$, $E_d < 0$ is the donor level position counted from $E_C$.

Fermi level position $E_F(T)$ should be determined self-consistently from the electro-neutrality condition $N_{d0}^+ - n_0 = 0$ valid in the single-domain region of ferroelectric, where potential vanishes and strains tend to the spontaneous values. Elementary calculations made in Boltzmann approximation lead to the expressions:

$$E_F(T) = \frac{E_d + E_C}{2} - \frac{k_B T}{2} \ln\left(\frac{\sqrt{\pi m_n^3 k_B^3 T^3}}{\sqrt{2}\pi^2 \hbar^3 N_d^0}\right), \quad \text{(S.8a)}$$

$$n_0(T) = \sqrt{\frac{N_d^0 \sqrt{\pi m_n^3 k_B^3 T^3}}{\sqrt{2}\pi^2 \hbar^3}} \exp\left(\frac{E_d - E_C}{2 k_B T}\right). \quad \text{(S.8b)}$$

Neel component of the polarization, that is perpendicular to the DW plane, leads to the depolarization field $\tilde{E}_1(\tilde{x}_1)$ across the wall, which profile is almost anti-phase to $\tilde{P}_1(\tilde{x}_1)$ (compare **Fig. S3a** with **S3b**). Depolarization field $\tilde{E}_1(\tilde{x}_1)$ induces potential barrier (or well) $\varphi(\tilde{x}_1)$ along the wall (**Fig. S3c**). One can see from the figure that the difference between full-scale calculations with account of semi conductive properties and the ones calculated in dielectric limit is negligibly small.

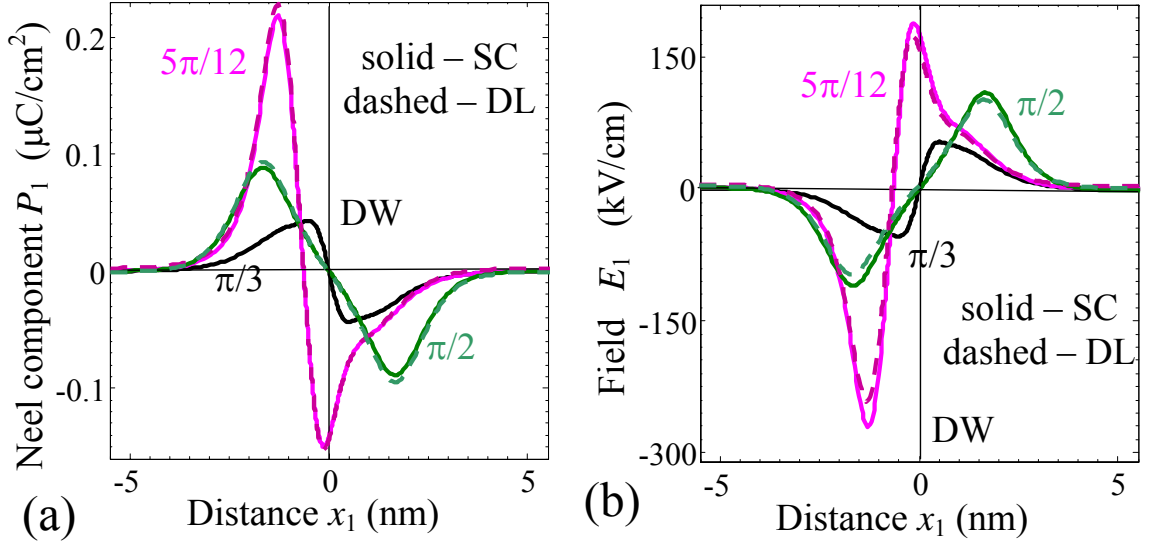

**Figure S3.** Profiles of Neel $\widetilde{P}_1(\widetilde{x}_1)$ polarization component (a), depolarization field $\widetilde{E}_1(\widetilde{x}_1)$ (b) calculated across the DW for rotation angles $\alpha = \pi/3, 5\pi/12, \pi/2$ (specified near the curves), temperature 200 K, flexoelectric coefficients $F_{11}= 2.46$, $F_{12}=0.48$, $F_{44}=0.05$ in $10^{-11}C^{-1}m^3$ and BaTiO$_3$ parameters listed in the **Table S1**. Solid curves correspond to full-scale calculations with account of semiconducting properties (SC): $n_0 = 3\times10^{22} m^{-3}$, parameters $E_d - E_C = -0.1$ eV, $N_d^0 = 10^{23} m^{-3}$, $m_n = 0.3 m_e$; dashed curves are calculated in dielectric limit $n_0 = 0$ (DL).

Current-AFM contrast strongly increases with the temperature decrease due to the increase of the potential jump ratio $e|\varphi(0)|/k_B T$ at the wall with temperature decrease (shown in **Fig. S4a**. However the concentration $n_0$ strongly decreases with temperature decrease as shown in **Fig. S4b.**

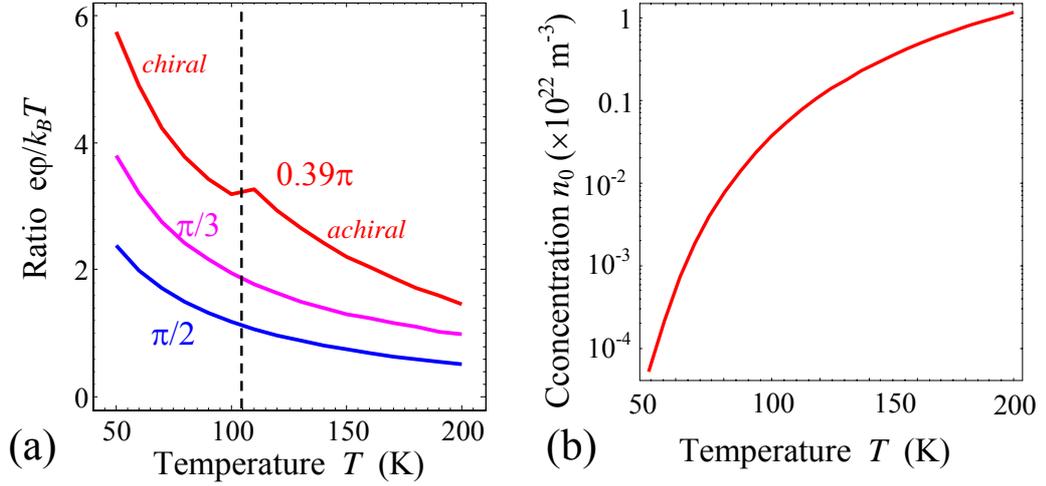

**Figure S4. (a)** Temperature dependence of the (a) potential jump ratio $e\varphi/k_B T$ at the DW (b), relative electron density of a single domain region (c) in rhombohedral ferroelectric phase of BaTiO$_3$ calculated for different rotation angles $\alpha = \pi/2, \pi/3, 0.39\pi$ specified near the curves. Plot (b) was calculated using Eq.(7) and parameters $E_d - E_C = -0.1$ eV, $N_d^0 = 10^{23}$ m$^{-3}$, $m_n = 0.3 m_e$.

## Appendix S6. Properties of nominally uncharged 180-degree domain wall in rhombohedral BaTiO$_3$

Nominally uncharged 180-degree domain wall properties in rhombohedral BaTiO$_3$ in dependence on the wall rotation angle $\alpha$ with respect to the {111} crystallographic orientation are summarized in the **Table S3.**

**Table S3 Properties of nominally uncharged 180-degree domain wall in rhombohedral BaTiO$_3$**

| Domain wall properties | Without flexoelectric coupling | With flexoelectric coupling coefficients |
|---|---|---|
| **Polarization vector structure** | Periodic with period $\pi/3$ and so the wall is superposed with itself after rotation on $\pi$.<br>Neel component is absent for $\alpha = m\pi/3$ (*m* is integer). | Periodic with period $2\pi/3$, so despite the wall is geometrically superposed with itself after rotation on $\pi$, the energy and structure of these walls are different. Flexoelectric coupling breaks the wall $\pi$-rotation symmetry and gives birth to "positive" and "negative" domain walls (which have different signs of Ising component gradient).<br>Neel component is nonzero for all $\alpha$. |
|  | Stable achiral Ising-odd-Bloch wall corresponds to $\alpha = m\pi/3$.<br>Achiral Ising-Bloch-Neel wall is energetically preferable in the $\pi/12$-vicinity of $\alpha = m\pi/3$.<br>Metastable chiral Ising-even-Bloch Neel wall corresponds to $\alpha = (1+2m)\pi/6$.<br>Mixed chiral Ising-Bloch-Neel walls are energetically preferable in the $\pi/12$-vicinity of $\alpha = (1+2m)\pi/6$. ||
| **Polarization** | Bloch component maximal value ~(10-20)$\mu$C/cm$^2$ is comparable with $P_S \sim 27$ $\mu$C/cm$^2$. ||
| **maximum value** | Neel component maximal value ~(0.1-0.2) µC/cm$^2$ is much smaller. | |
| | Bloch component is maximal at $\alpha = \pi/6 + m\pi/3$ and minimal at $\alpha = \pi/6 \pm \pi/12 + m\pi/3$. Neel component is maximal at $\alpha = \pi/6 \pm \pi/12 + m\pi/3$ and identically zero (global minimum) at $\alpha = m\pi/3$. | Bloch component absolute maximum corresponds to $\alpha = \pi/2 + 2m\pi/3$. Local maximums correspond to $\alpha = \pi/6 + 2m\pi/3$. Neel component absolute maximum corresponds to $\alpha \approx \pi/2 \pm \pi/12 + 2m\pi/3$; local minimums correspond to $\alpha \approx m\pi/12$ |
| **Chirality** | Odd-type Ising-Bloch-Neel walls are achiral. Domain walls that can be switched from the left-handed state to the right-handed one by electric field | |
| **Intrinsic energy** | Periodic with period $\pi/3$, "positive" and "negative" domain walls are indistinguishable. Left- and right-handed walls have the same energy. Energy is minimal at $\alpha = m\pi/3$ and maximal at $\alpha = \pi/6 + m\pi/3$. | Periodic with period $2\pi/3$, since "positive" and "negative" domain walls have slightly different energies. Energy is minimal at $\alpha = m\pi/3$. Global maximums correspond to $\alpha = \pi/2 + 2m\pi/3$. Local maximums correspond to $\alpha = \pi/6 + 2m\pi/3$. |
| **c-AFM contrast** | Walls with maximal Neel component ($\alpha = \pi/6 \pm \pi/12 + m\pi/3$) are the most conductive. Walls with zero Neel component ($\alpha = m\pi/3$) has the same conductivity as the single-domain bulk. | Walls with maximal Neel component ($\alpha \approx \pi/2 \pm \pi/12 + 2m\pi/3$) are the most conductive. Walls with minimal Neel component ($\alpha \approx m\pi/12$) are less conductive, but still more conductive than the single-domain bulk. |
| | Neel component charges the "nominally uncharged" 180-degree wall. Depending the temperature and flexoelectric coupling strength the wall static conductivity becomes at least one order higher than in the single-domain region, creating the contrast pronounced and easily detectable by c-AFM. | |